\documentclass{PoS}

\usepackage{transparent}
\usepackage{amsmath}
\usepackage{amssymb}
\usepackage{bm}
\usepackage[table,x11names]{xcolor}
\usepackage{graphicx}
\usepackage{gensymb}
\usepackage[font=footnotesize,labelfont=bf]{caption}
\usepackage{subcaption}

\title{Techniques for Measuring Galactic Diffuse Emission Flux and their Preliminary Results in Confused Regions}

\ShortTitle{Techniques for Measuring Galactic Diffuse Emission}

\author{\speaker{Chang Dong Rho}\\
        University of Rochester\\
        E-mail: \email{crho2@ur.rochester.edu}}

\author{Hugo Ayala\\
       Pennsylvania State University\\
       E-mail: \email{hayalaso@mtu.edu}}
       
\author{Hao Zhou\\
       Los Alamos National Laboratory\\
       E-mail: \email{hayalaso@mtu.edu}}

\author{For the HAWC Collaboration\\
		For a complete author list, see www.hawc-observatory.org/collaboration/icrc2017.php.}

\abstract{Galactic diffuse emission has provided us with evidence for cosmic ray acceleration throughout the Galaxy and the background for searches for physics beyond the Standard Model. However, only the very limited measurements of the diffuse flux are available in TeV $\gamma$ rays. The High Altitude Water Cherenkov (HAWC) Observatory is well-suited for observing the diffuse emission of very high energy with its unbiased, wide field of view (2 sr). Using data from HAWC, we present techniques for measuring the diffuse flux and show preliminary results.}

\FullConference{35th International Cosmic Ray Conference --- ICRC2017\\
		10--20 July, 2017\\
		Bexco, Busan, Korea}

\begin{document}

\section{Introduction}\label{sec:intro}
The Galactic diffuse flux we observe through high-energy Galactic $\gamma$ rays is comprised of unresolved point sources and "true" diffuse emission. In the GeV to TeV range, there are three main contributions to the diffuse emission, namely, inverse Compton scattering, $\pi^{0}$ decay and bremsstrahlung. In brief, the production mechanisms involve:
\begin{itemize}
  \item Inverse Compton scattering of cosmic microwave background, far infrared and optical photons by cosmic ray electrons.
  \item Interaction of cosmic rays with the interstellar medium which forms neutral pions that decay into gammas ($\pi^{0}\rightarrow\gamma\gamma$).
  \item Deceleration of cosmic rays when propagating through the interstellar medium, hence the production of $\gamma$ rays through bremsstrahlung.
\end{itemize}

Unlike at GeV energies, at TeV, the diffuse emission may seem to be relatively less important than the contributions from point sources \cite{abe17}. Hence, analyses of diffuse $\gamma$-ray emission at TeV energies are still sparse. However, diffuse emission in the inner Galaxy may be harder in terms of spectrum than in the outer parts of the Galaxy \cite{gag17}, making it more significant than anticipated in the TeV regime.

The Galactic diffuse flux is of intrinsic interest at all wavelengths and is a key tracer of cosmic ray acceleration, diffusion, and propagation in the Galaxy. While the measurements of charged cosmic rays such as elemental abundances and particle spectra can depend strongly on the Galactic magnetic field and local features in the distribution of Galactic objects, the $\gamma$ rays provide a global picture of the cosmic-ray environment in the Milky Way. $\gamma$ rays can, therefore, be used to test models of particle diffusion and provide complementary information about cosmic-ray sources more effectively than the observed cosmic rays on Earth.

Although our understanding of Galactic diffuse emission has progressed extensively for the past few years, we still face many unanswered problems such as the distribution of the interstellar gas, the models of cosmic ray diffusion and $\gamma$-ray diffusivity. Also, having a deeper understanding of the diffuse emission can assist in separating unresolved sources from the "true" diffuse flux, aiding identification of unknown $\gamma$-ray sources. The knowledge of the diffuse flux is especially important in confused regions in proximity of the Galactic Plane. 

\section{HAWC}\label{sec:hawc}
The High Altitude Water Cherenkov (HAWC) Observatory is a ground array located at latitude of 19\degree N and at an altitude of 4,100 meters in Sierra Negra, Mexico. HAWC consists of 300 water Cherenkov detectors (WCDs) covering a large effective area of 22,000 m$^2$. Of the 300 deployed tanks, 294 have been instrumented \cite{hawccrabpaper}. Each WCD has a light-tight polypropylene bladder filled with 200 kL of purified water. The bladder is encased in a steel tank. At the bottom of each WCD there are three 8-inch Hamamatsu R5912 photomultiplier tubes (PMTs) oriented in an equilateral triangle and one 10-inch R7081-HQE PMT anchored at the center.

By combining the location and the time of each PMT triggered by an air shower, the core position and the angle at which the primary particle has generated the air shower is reconstructed to locate and identify the primary particle type. Simple topological cuts are applied to discriminate the air showers produced by hadronic cosmic rays from $\gamma$-ray air showers. For a detailed explanation of the event reconstruction, see \cite{smi15}.

The air shower trigger rate of HAWC is approximately 25 kHz, more than 99.9\% of which are events originating from cosmic rays. At HAWC's altitude, a vertical shower from a 1 TeV photon will have about 7\% of the original photon energy. This energy increases to around 28\% at 100 TeV \cite{hawccrabpaper}. The main source of background to $\gamma$-ray observation is the hadronic cosmic-rays. Therefore, individual $\gamma$-ray-induced air showers are distinguished from cosmic-ray showers using their topology. HAWC has a duty cycle $>95\%$ and a wide, unbiased field of view of ~2 sr. As such, HAWC is well-suited to study spatially extended structures, making it an excellent detector to study the Galactic TeV $\gamma$-ray diffuse flux.

\section{Simple Gaussian Model}\label{sec:sgm}
Since the definition of diffuse emission requires measurement of flux not attributable to known TeV $\gamma$-ray sources, the analysis of the diffuse flux using data from HAWC can be achieved using two different methods: assessment of residual flux after subtracting all the known TeV $\gamma$-ray sources \cite{hao}; or, simultaneously fitting the diffuse emission and the known TeV sources in the region of interest based on an estimated diffuse model. The latter is discussed in this paper. In both cases, unresolved sources pose a considerable problem. At TeV, however, the $\gamma$-ray flux from resolved sources make up most of the total observed Galactic $\gamma$-ray emission \cite{abe17,abr14}. Therefore, determining the most accurate Galactic diffuse model is essential. In this section, we have assumed a latitudinal profile of a simple Gaussian function \cite{abr14} to describe the Galactic diffuse emission flux, 
\begin{equation}
F=Ke^{\frac{-b^{2}}{2\sigma_b^{2}}},
\label{eqn:gaussian}
\end{equation}
where $K$ is flux normalization, $b$ is galactic latitude at a specific pixel and $\sigma_b$ is Gaussian width. Equation~\ref{eqn:gaussian} is used to model the Galactic diffuse flux while freeing the flux normalization and the Gaussian width during likelihood maximization.

\begin{figure}[!htb]
  \centering
  \begin{subfigure}[b]{0.45\textwidth}
  \includegraphics[width=\linewidth]{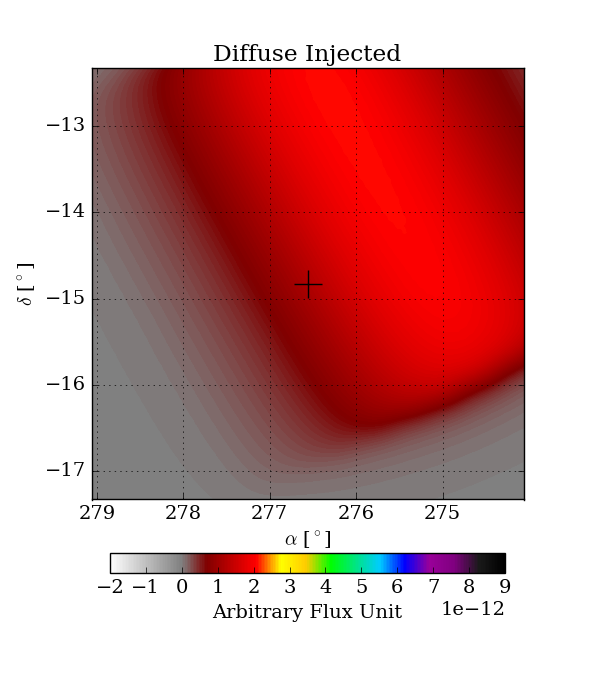}
  \end{subfigure}
  \begin{subfigure}[b]{0.45\textwidth}
  \includegraphics[width=\linewidth]{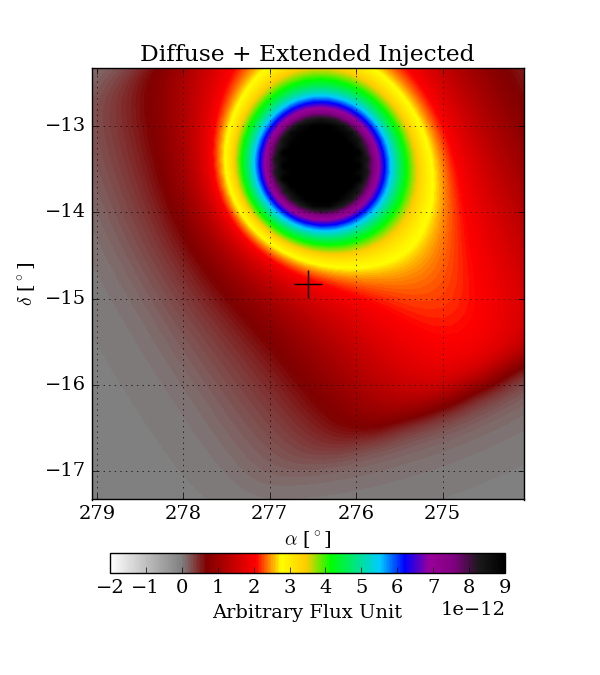}
  \end{subfigure}
  \begin{subfigure}[b]{0.45\textwidth}
  \includegraphics[width=\linewidth]{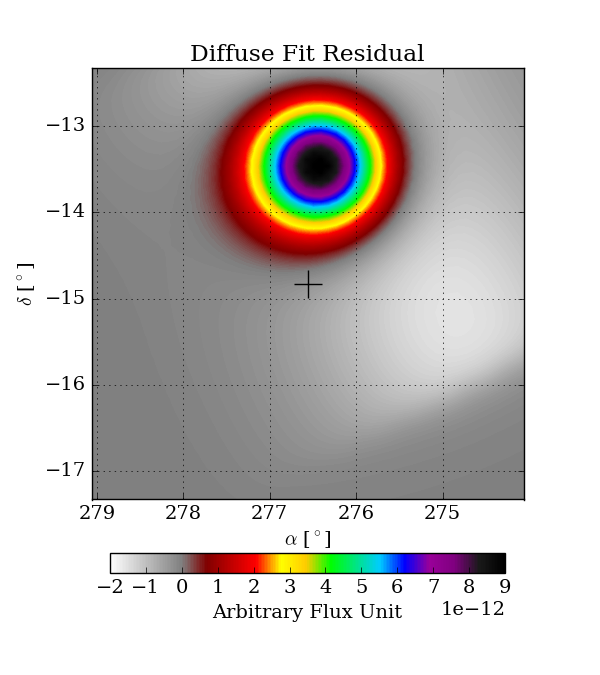}
  \end{subfigure}
  \begin{subfigure}[b]{0.45\textwidth}
  \includegraphics[width=\linewidth]{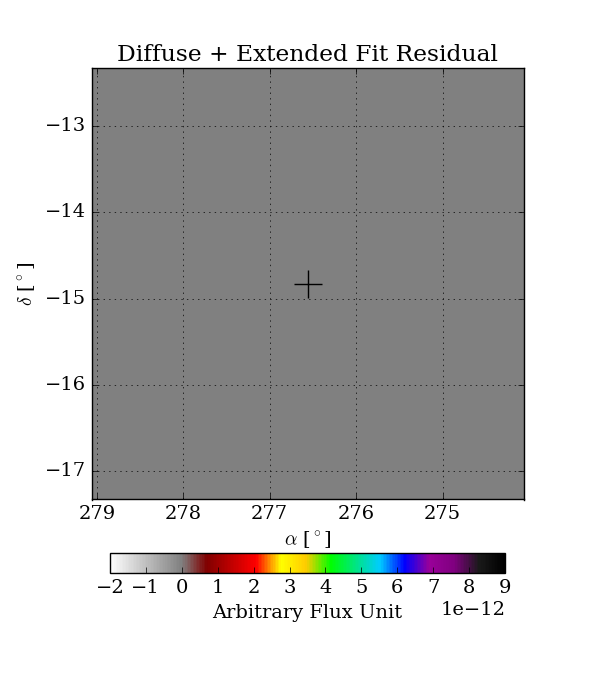}
  \end{subfigure}
  \put(-250,-10){\huge\transparent{0.2}\color{red}{Preliminary}}
  \caption{Top left: A map showing simulated diffuse source injected. Top right: A map after the injection of simulated diffuse source and a simulated extended source. Bottom left: After fitting and subtracting only the diffuse source from the top right plot. Bottom Right: After multi-source fitting and subtracting the diffuse source and the extended source from the top right plot.}
  \label{fig:stimls5039}
\end{figure}

Before applying the diffuse model to data, an extended source and a Gaussian diffuse source models were injected into a pseudo map to simulate a confused region to test simultaneous fits. This is displayed in Figure~\ref{fig:stimls5039} where the color bar shows an arbitrary flux unit. The first plot on the top left shows only the injected Gaussian diffuse source and the plot on the top right shows the injected diffuse and extended sources. The bottom left is the residual plot after fitting and subtracting the Gaussian diffuse source from the top right plot. There are regions where over subtractions are visible because only the diffuse source was fitted and subtracted. The plot on the bottom right shows the residual map after subtracting the diffuse and extended source models fitted simultaneously. The application of multi-source fit displays an improved fit result of the diffuse source when another source exists.

\begin{figure}[!htb]
  \centering
  \begin{subfigure}[b]{0.45\textwidth}
  \includegraphics[width=\linewidth]{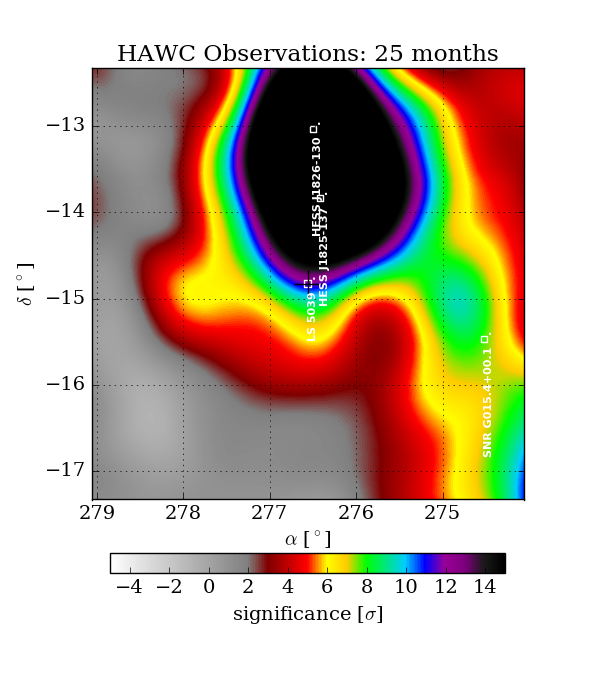}
  \end{subfigure}
  \begin{subfigure}[b]{0.45\textwidth}
  \includegraphics[width=\linewidth]{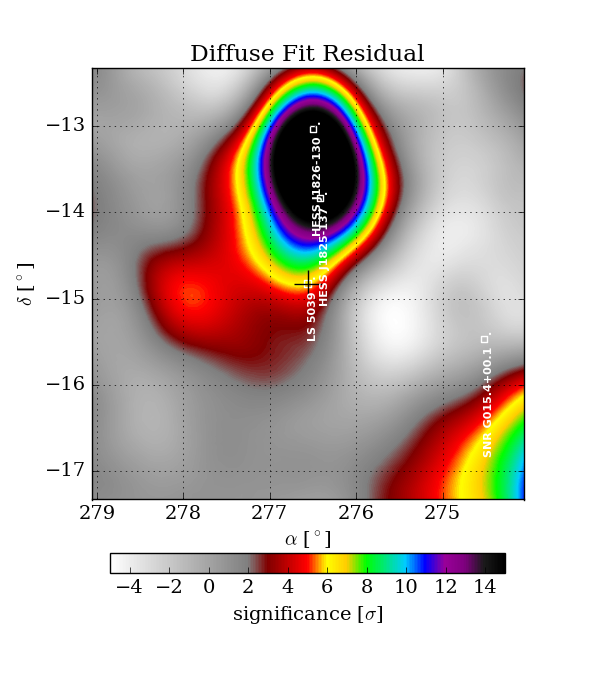}
  \end{subfigure}
  \begin{subfigure}[b]{0.45\textwidth}
  \includegraphics[width=\linewidth]{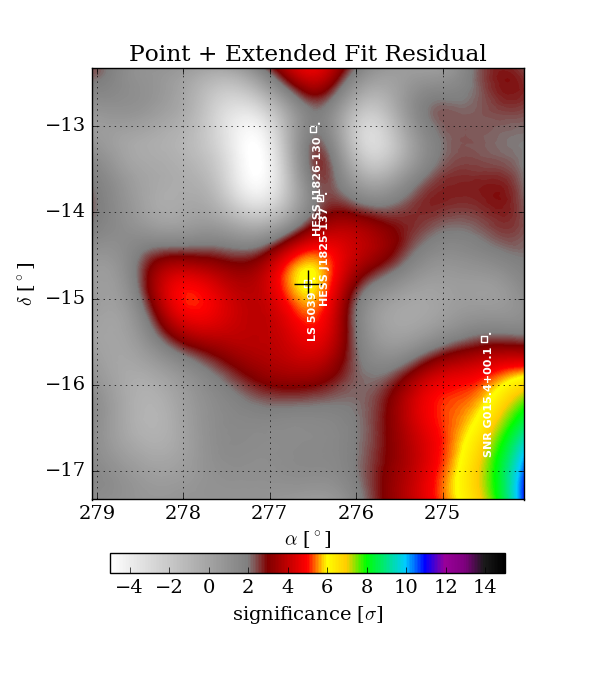}
  \end{subfigure}
  \begin{subfigure}[b]{0.45\textwidth}
  \includegraphics[width=\linewidth]{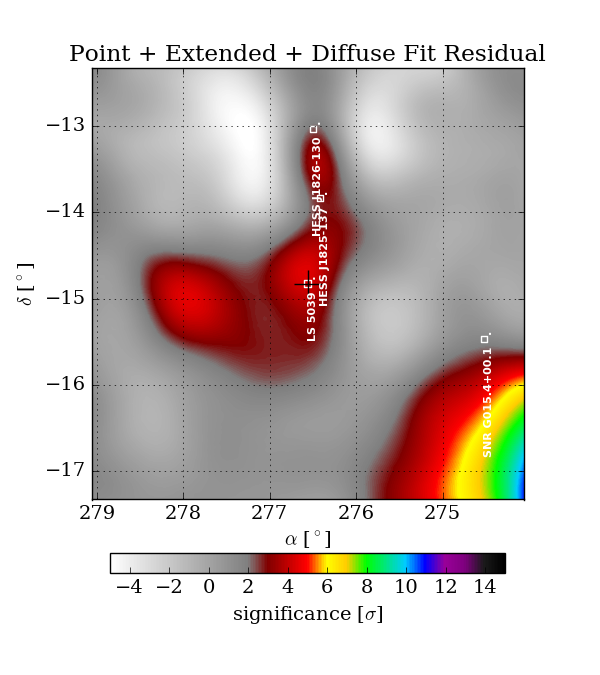}
  \end{subfigure}
  \put(-250,0){\huge\transparent{0.2}\color{red}{Preliminary}}
  \caption{Top left: Significance map of LS 5039 region. Top right: Residual map after fitting and subtracting the Gaussian diffuse flux. Bottom left: Residual map after fitting and subtracting the 3 known sources (2HWC J1812-126, 2HWC J1819-150 and 2HWC J1825-134) in the region of interest. Bottom right: After simultaneously fitting and subtracting the Gaussian diffuse flux and the known sources.}
  \label{fig:ls5039stuff}
\end{figure}

For the analysis with 25 months of HAWC data, the region of interest was set between $-3\degree$ and $3\degree$ in Galactic latitude and between $15\degree$ and $21\degree$ in Galactic longitude. This was to select a source confused region on the Galactic Plane. Table 1 shows the Test Statistics \cite{you15} (TS) of the different fits attempted in the same region of interest. With multi-source fits, the TS increases, meaning that a better fit is being performed with added source models. Also, the residual plots, after subtracting the resolved sources, are shown in Figure~\ref{fig:ls5039stuff} in the form of significance maps \cite{you15}. The known sources in the defined region include: 2HWC J1812-126; 2HWC J1819-150; and 2HWC J1825-134 \cite{abe17}. The result from Figure~\ref{fig:ls5039stuff} displays how HAWC is capable of improving fit results by applying a hypothetical diffuse model; namely by comparing bottom left and bottom right plots in Figure~\ref{fig:ls5039stuff}. This is done to demonstrate the importance of diffuse emission when studying a source confused region, rather than to compute the correct diffuse emission model using data. Once we become more confident with the simple Gaussian model, we plan to measure the diffuse emission in HAWC data using sliding window scans. The strips will have small overlaps to reduce edge effects.

\begin{table}[ht]
\centering
\footnotesize
\begin{tabular}{l|r|r}
\bf Fit Description &
$\mathbf{Test~Statistics~(TS)}$ & 
\bf \#~Free~Parameters\\
\hline
Diffuse Only        &  1102.6  &  2 \\
Multi-source (Extd + PS)        &  1702.3  &  3 \\
Multi-source (Diffuse + Extd + PS)        &  1826.0  &  5 \\
\end{tabular}
\caption{Fits results of LS 5039 region using 25 month dataset. First row shows the result of the diffuse flux fit only. Second row shows the result of multi-source fitting the three known sources in the region. Third row shows the result of multi-source fitting the three known sources and the diffuse flux.}
\label{table:tss433}
\end{table}

\pagebreak
\section{Template Model}\label{sec:galprop}
Another way to obtain the diffuse emission flux information is by using models from simulations that take into account measurements of cosmic rays and distribution of gas and interstellar radiation fields in the Galaxy. Hence, measurement of diffuse emission using existing template models with data from HAWC is also under development. 

\begin{figure}[!htb]
  \centering
  \begin{subfigure}[b]{0.45\textwidth}
  \includegraphics[width=\linewidth]{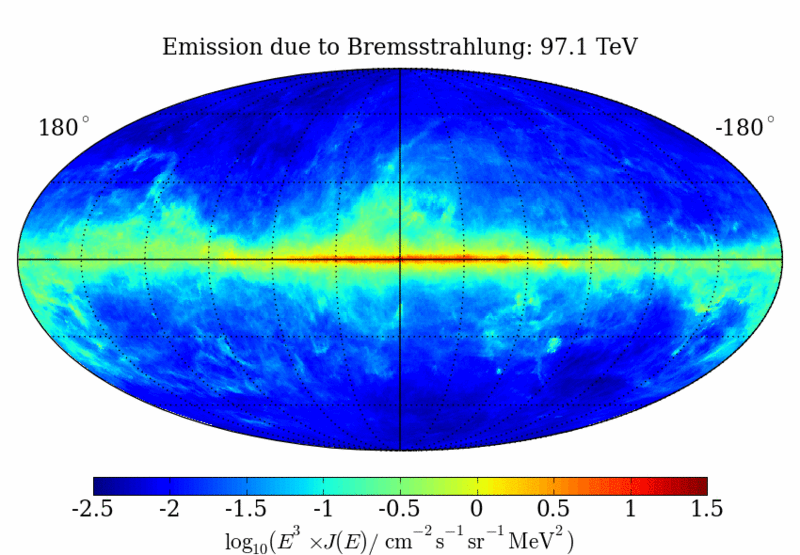}
  \end{subfigure}
  \begin{subfigure}[b]{0.45\textwidth}
  \includegraphics[width=\linewidth]{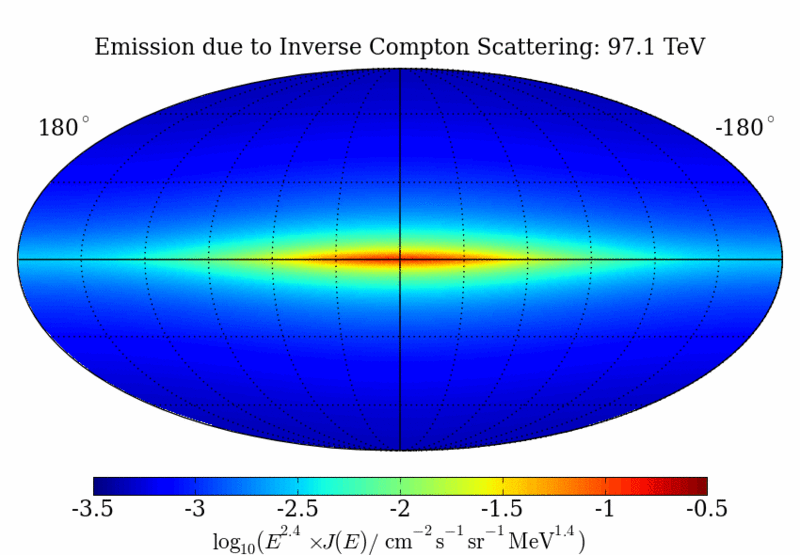}
  \end{subfigure}
  \begin{subfigure}[b]{0.45\textwidth}
  \includegraphics[width=\linewidth]{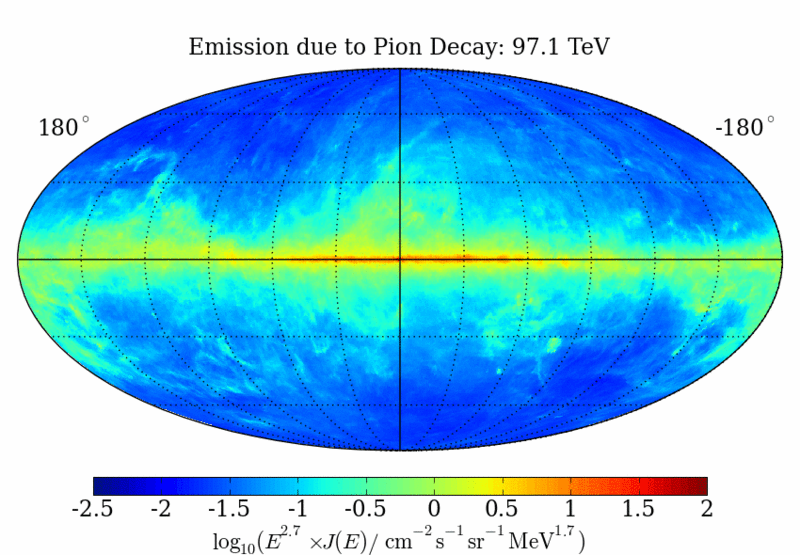}
  \end{subfigure}
  \caption{Display of GALPROP model that predicts diffuse emission at TeV energies. Three contributions - bremsstrahlung, inverse Compton scattering and neutral pion decay - are available. The files are 3D: Galactic longitude, Galactic latitude and energy. The bin size is 0.5$\degree$ and the maps contain the whole sky \cite{str15}.}
  \label{fig:galprop}
\end{figure}

GALPROP is a cosmic ray propagation code that specializes in the analysis of diffuse $\gamma$ rays and cosmic rays. The GALPROP project began in 1996 \cite{mos98} and has been under a continuous development since. The code was made public in 1998 and the analysis discussed here will use version 54, released in 2011 \cite{vla11}. The GALPROP code uses various inputs such as direct measurements of primary and secondary cosmic ray nuclei, electrons and positrons, $\gamma$ rays, synchrotron radiation and so on \cite{mos11}. These are all interlinked to explain multiwavelength and multi-messenger observations \cite{str98}. Three contributions (bremsstrahlung, inverse Compton scattering and neutral pion decay) available for the GALPROP template at 97.1 TeV are presented in Figure~\ref{fig:galprop} \cite{str15}.

DRAGON is another code that solves the diffusion equation of interstellar cosmic rays, including inhomogeneous and anisotropic diffusion \cite{dragon}. DRAGON is based on GALPROP version 50, but it allows and solves the transport equation in 3D, compared to GALPROP which uses a 2D model.

The idea is to use a template model, such as GALPROP or DRAGON, that predicts diffuse emission at TeV energies and implement them for the HAWC $\gamma$-ray data challenge. The goal is to use them as a template in HAWC's own likelihood minimization program to produce diffuse flux measurements on TeV $\gamma$ rays. The template is conformed by 3-axes: one energy axis; and two spatial axes given in Galactic coordinates. The template will give the pixel by pixel flux information of the region of interest. The flux in each map is obtained by using the interpolation method, \textit{RegularGridInterpolator}, in the SciPy library \cite{scipy}.

The calculated flux will be convolved with the detector response of the HAWC detector. This will give the number of expected events that will be used in the likelihood analysis. In order to fit the diffuse emission with the template, the flux normalization factor is set free. The likelihood calculation is performed by the Multi-Mission Maximum Likelihood (3ML) framework \cite{3ml} and the Likelihood Fitting Framework (LiFF) \cite{liff}.

\section{Conclusion}\label{sec:conclusion}
HAWC is capable of observing and measuring TeV $\gamma$ rays produced by Galactic diffuse emissions. Having a good comprehension of the diffuse emission in the TeV regime is important. Moreover, analysis of pointlike and extended sources near the Galactic Plane require a good understanding of the diffuse emission. Therefore, different techniques are being developed within the HAWC collaboration and this paper introduces some of these at work. The simple Gaussian method is already being used for multi-source fits and an example of this in a confused region has been presented. Also, preliminary diffuse flux measurements such as longitudinal and latitudinal profiles are under study \cite{hao}. The template method, on the other hand, is expected to allow existing Galactic diffuse templates, based on cosmic-ray calculations from GALPROP / DRAGON, to produce maximum likelihood fits to the data with 3ML \cite{3ml}. Once available, the template method should also be compatible with multi-source fits.

\end{document}